\documentclass[conference]{IEEEtran}
\IEEEoverridecommandlockouts
\usepackage{cite}
\usepackage{amsmath,amssymb,amsfonts}
\usepackage{graphicx}
\usepackage{textcomp}
\usepackage{xcolor}
\usepackage{hyperref}

\def\BibTeX{{\rm B\kern-.05em{\sc i\kern-.025em b}\kern-.08em
    T\kern-.1667em\lower.7ex\hbox{E}\kern-.125emX}}
\begin{document}

\title{Two-dimensional structure functions to characterize convective rolls in the marine atmospheric boundary layer from Sentinel-1 SAR images. \\
\thanks{This work was supported by the French National Research Agency (ANR-21-CE46-0011-01), within the program ”Appel \`a projets g\'en\'erique 2021”.}
}

\author{\IEEEauthorblockN{Carlos Granero-Belinchon\IEEEauthorrefmark{1}\IEEEauthorrefmark{4}, St\'ephane G. Roux\IEEEauthorrefmark{2}, Nicolas B. Garnier\IEEEauthorrefmark{2}, Pierre Tandeo\IEEEauthorrefmark{1}, \\ Bertrand Chapron\IEEEauthorrefmark{3} and Alexis Mouche\IEEEauthorrefmark{3}}
\IEEEauthorblockA{\IEEEauthorrefmark{1}\textit{Mathematical and Electrical Engineering Department}, 
\textit{IMT Atlantique, Lab-STICC, UMR CNRS 6285},
F-29238 Brest, France}
\IEEEauthorblockA{\IEEEauthorrefmark{2}\textit{Laboratoire de Physique},
\textit{Univ Lyon, Ens de Lyon, CNRS UMR 5672,}
F-69342 Lyon, France}
\IEEEauthorblockA{\IEEEauthorrefmark{3}\textit{Laboratoire d'Oc\'eanographie Physique et Spatiale},
\textit{Ifremer, Univ. Brest, CNRS, IRD, IUEM},
Plouzan\'e, France}
\IEEEauthorblockA{\IEEEauthorrefmark{4}Corresponding author: carlos.granero-belinchon@imt-atlantique.fr}}

\maketitle

\begin{abstract}
We study the shape of convective rolls in the Marine Atmospheric Boundary Layer from Synthetic Aperture Radar images of the ocean. We propose a multiscale analysis with structure functions which allow an easy generalization to analyse high-order statistics and so to finely describe the shape of the rolls. The two main results are : 1) second order structure function characterizes the size and direction of rolls just like correlation or power spectrum do, 2) high order statistics can be studied with skewness and Flatness which characterize the asymmetry and intermittency of rolls respectively. From the best of our knowledge, this is the first time that the asymmetry and intermittency of rolls is shown from radar images of the ocean surface.
\end{abstract}


\section{Introduction}

Depending on the surface layer stratification, rolls approximatively aligned with the mean flow can develop in the Marine Atmospheric Boundary Layer (MABL) into Organised Large Eddies (OLE). Although, this secondary circulation plays a key role in turbulent heat and momentum fluxes~\cite{Zhang2008}, MABL fluxes parameterization and modelling is still an active field of research for both the numerical models~\cite{Sandu2013} and Large Eddies simutations~\cite{Saggiorato2020}. One of the reasons is certainly that ground truth for collocated wind, temperature and humidity profile observations and for documenting OLE characteristics in open ocean are rare. As summarized by~\cite{Etling1993}, the main known characteristics are the following: vertical  extent: 1-2 km, wavelength: 2-20 km, aspect ratio: 2-15, downstream extent: 10-1000 km, orientation of roll axis to mean wind direction: -20° to +30°; and lifetime: 1-72 h. Yet, more advanced information such as their frequency of occurrence, strength or formation remains to be documented. For a more complete review on OLE from the observational, theoretical and numerical point of view, the reader can refer to review papers from~\cite{Etling1993} and~\cite{Young2002}.

C-Band high-resolution radar (or SAR for synthetic aperture radar) is the only spaceborne instrument able to probe at very high spatial resolution ocean sea surface day and night, regardless of the cloud coverage, with pixel resolution of few meters in swaths of several hundred kilometers. At C-band  (about 5~cm), the backscattered signal from the ocean is primary sensitive to capillary and small gravity ocean surface waves which are very sensitive to the local surface stress allowing for the SAR to trace the updraft and downdraft associated to horizontal wind field convergence and divergence areas in between the rolls. As a consequence, several studies have been conducted to report on the rolls signature on SAR images~\cite{Alpers1994} and document the  spatial evolution of convection for particular events such as strong cold outbreak interacting with the Gulf Stream north wall~\cite{Babin2003}. Most of them focused on coastal areas where large images of several hundreds of kilometers were acquired with previous SAR missions Envisat and Radarsat. But, very few take benefit of the acquisition mode used in open ocean : the so-called wave mode. In fact, to our knowledge, the only studies based on wave mode have been published by~\cite{Wang2020} for OLE detection and~\cite{Wang2019} for OLE documentation (orientation and wind speed modulation of $\pm$ 7 m/s). In particular, these two recent studies rely on the new capabilities of Sentinel-1 European SAR mission that provides images of 20 by 20 km over open ocean whereas the former European SAR (Envisat) was only able to provides scenes of 8 by 10 km, much less adapted to the OLE wavelength. Moreover, Wave Mode acquisition have the highest resolution: about 5~m.

Until now, previous works only used second order statistics to characterize size and direction of convective rolls directly from radar imagery of the atmosphere~\cite{Lohou1998} or from SAR imagery of the ocean~\cite{Wang2020}.
However, higher order statistics have been already used to characterize turbulence in 1-dimensional signals of wind velocity in the MABL~\cite{Atta1970}.
Thus, we propose here to use higher order statistics to characterize more subtly the shape of convective rolls. To do so, we compute the 2-dimensional skewness $\mathcal{S}$ and flatness $\mathcal{F}$ of spatial increments and examine their evolutions across length-scales~\cite{Frisch1995}, which allows us to describe the asymmetry and intermittency of convective rolls. To the best of our knowledge, it is the first time that such an analysis is performed on SAR images.

\section{Theoretical relations}

\subsection{Two-dimensional direction-dependent structure functions}

The n$^{th}$ order structure function $S_{n}$ of a two-dimensional field $F(x,y)$ can be defined as:

\begin{equation}
S^{l_x,l_y}_{n}(F) = \left\langle \left( F(r_x+l_x,r_y+l_y) - F(r_x,r_y) \right)^{n} \right\rangle 
\end{equation}

\noindent where $(r_x,r_y)$ denotes a spatial position, and so $l_x$ and $l_y$ are the separation distances along each dimension of the field. We then note $\delta_{l_x,l_y}F = F(r_x+l_x,r_y+l_y) - F(r_x,r_y)$, the spatial increment of the two-dimensional field.

In this paper we focus on $S_{2}^{l_x,l_y}$, $S_{3}^{l_x,l_y}$ and $S_{4}^{l_x,l_y}$ that provide respectively a characterization of the variance, asymmetry and tails prominence of the statistical distribution of the increments of the field at scales $(l_x,l_y)$. To avoid the impact of $S_{2}^{l_x,l_y}$ variations on the characterization of $S_{3}^{l_x,l_y}$ and $S_{4}^{l_x,l_y}$, the Skewness $\mathcal{S}$ and Flatness $\mathcal{F}$ factors across scales are defined. They correspond to $S_{3}^{l_x,l_y}$ and $S_{4}^{l_x,l_y}$ on centered and standardized increments.

\noindent If $\mathcal{S}=0$ then the distribution is symmetrical, while $\mathcal{S}<0$ implies left-tailed and $\mathcal{S}>0$ right-tailed distributions. On the other hand, the flatness of a Gaussian distribution is $\mathcal{F}=3$, while $\mathcal{F}<3$ means less prominent tails and $\mathcal{F}>3$ more prominent ones.

From a physical point of view, $S_{2}^{l_x,l_y}$ characterizes the distribution of energy across scales, while $\mathcal{S}^{l_x,l_y}$ and $\mathcal{F}^{l_x,l_y}$ allow us to describe intermittency which translates into a deformation of the shape of the distribution across scales~\cite{Frisch1995}. Thus for example turbulence is characterized by a Gaussian distribution at large scales and an increase, of skewness and flatness when the scale decreases~\cite{Kolmogorov1962, Obukhov1962, Frisch1995}.

Finally, a change of coordinates from cartesian to polar allows for analysis with respect to direction: $S_{2}^{l_x,l_y} \rightarrow S_{2}^{r,\theta}$, $\mathcal{S}^{l_x,l_y}\rightarrow\mathcal{S}^{r,\theta}$ and $\mathcal{F}^{l_x,l_y}\rightarrow\mathcal{F}^{r,\theta}$, with $r$ the radius and $\theta$ the angle dimension.

\subsection{Interpretation of structure functions analysis of OLE}

So, second order statistics such as power spectrum, correlation function or $S_{2}^{l_x,l_y}$, describe the size and direction of convective rolls~\cite{Lohou1998,Wang2020} without providing finer information nor on the shape of rolls neither on the evolution of this shape across scales. Thus from~\cite{Lohou1998}, the correlation function decreases the fastest along the direction perpendicular to the rolls $\theta_{\perp}$. Moreover, along this direction the correlation length, measured as the distance to the first minimum, indicates the size of the rolls. On the other hand from~\cite{Wang2020}, peaks of energy appear in the 2-dimensional power spectrum and the most energetic peaks indicate the size and direction of rolls. Since power spectrum and correlation (equivalently $S_{2}^{l_x,l_y}$) are simply related by the Fourier transform, both methodologies are equivalent. These second order statistics are frequently used in the characterization of ocean surface from remote sensing images~\cite{Tandeo2014}.

\begin{figure*}[htbp]
\centerline{\includegraphics[width=0.8\linewidth]{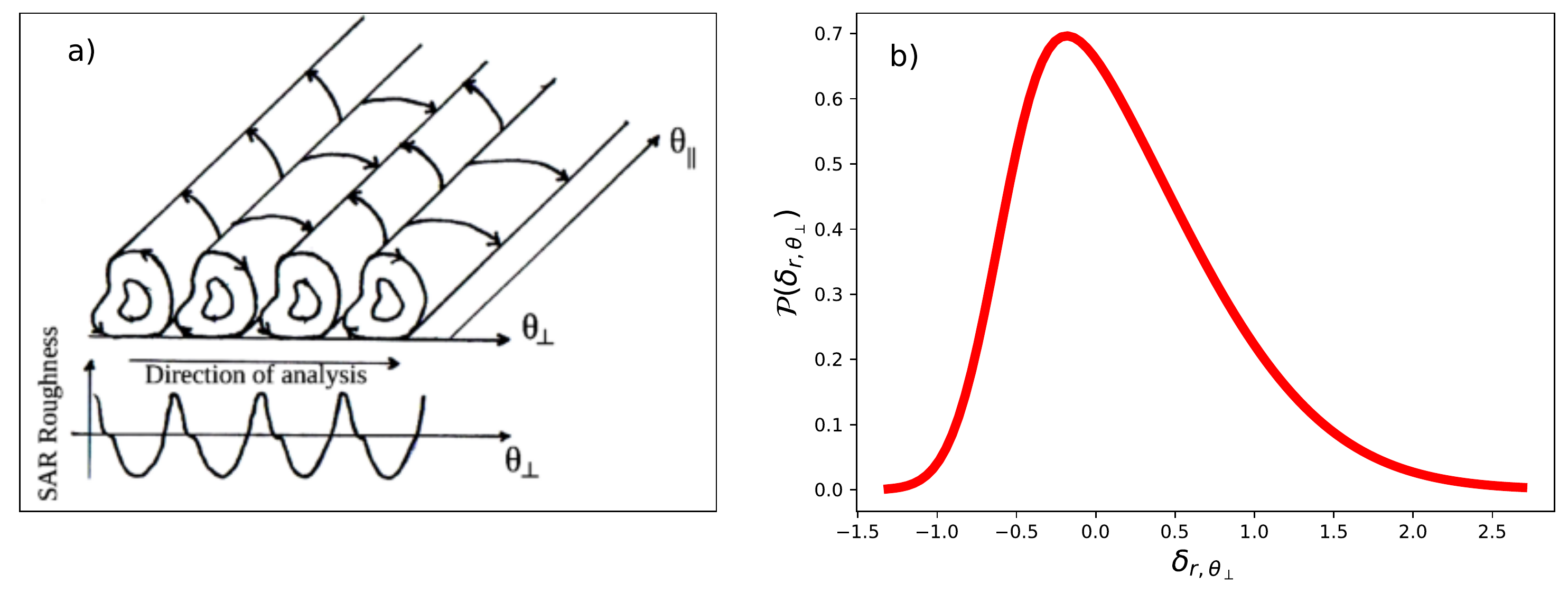}}
\caption{a) Outline of asymmetrical convective rolls and their SAR imprint. The direction of analysis corresponding to $\theta_{\perp}$ is indicated by the \textit{direction of analysis} arrow. b) Schematic distribution of the increments of the SAR roughness at a given scale $r$ smaller than the rolls wavelength and along the direction indicated by the \textit{direction of analysis} arrow in a). The positive skewed distribution of the increments points out the asymmetry of the rolls observed in a). In this example, roughness increases when leaving rolls are sharper than roughness decreases when entering into the rolls which leads to larger values of positive increments than those of negative ones and so to a positive skewed distribution of the increments.}
\label{figschema}
\end{figure*}

Contrary to second-order statistics, high order ones such as $\mathcal{S}^{l_x,l_y}$ and $\mathcal{F}^{l_x,l_y}$ characterize the shape of the rolls and its finer changes across scales. More precisely, non-zero values of $\mathcal{S}^{l_x,l_y}$ illustrate an asymmetry of the shape of the rolls along the direction pointed out by the couple $(l_x, l_y)$. Indeed, $\mathcal{S}^{l_x,l_y}$ characterizes the asymmetry of the distribution of the increments of size $(l_x, l_y)$, and so, characterizes the differences between rises and falls of the field along this direction and at this scale, see figure~\ref{figschema}.
We interpret low intensity areas of SAR fields as areas occupied by the rolls, while high intensity areas are interpreted as areas between rolls. Thus, $\mathcal{S}^{l_x,l_y}\neq0$ along $\theta_{\perp}$ implies that we enter and leave rolls in a different way (more or less smoothly/sharply) pointing out an asymmetry, see figure~\ref{figschema}.

\section{Sentinel-1 SAR Wave Mode dataset}\label{sec:data}

Nowadays, the largest database of C-band Synthetic Aperture Radar (SAR) images of the ocean is provided by the Sentinel-1 constellation mission. 
Sentinel-1A and Sentinel-1B twin SAR can operate in 4 different and exclusive acquisition modes. They differ in incidence angle range, swath width (spatial coverage), resolution and polarization
In this work we focus on the Wave (WV) acquisition mode, which provides images of $20 \times 20$ km (the smallest) with a spatial resolution of $5$ m (the highest). Then, this mode furnishes large enough images to study convective rolls (typical sizes of 2-20 km), while covering quite homogeneous areas and at very high spatial resolutions. 

\begin{figure}[htbp]
\centerline{\includegraphics[width=\linewidth]{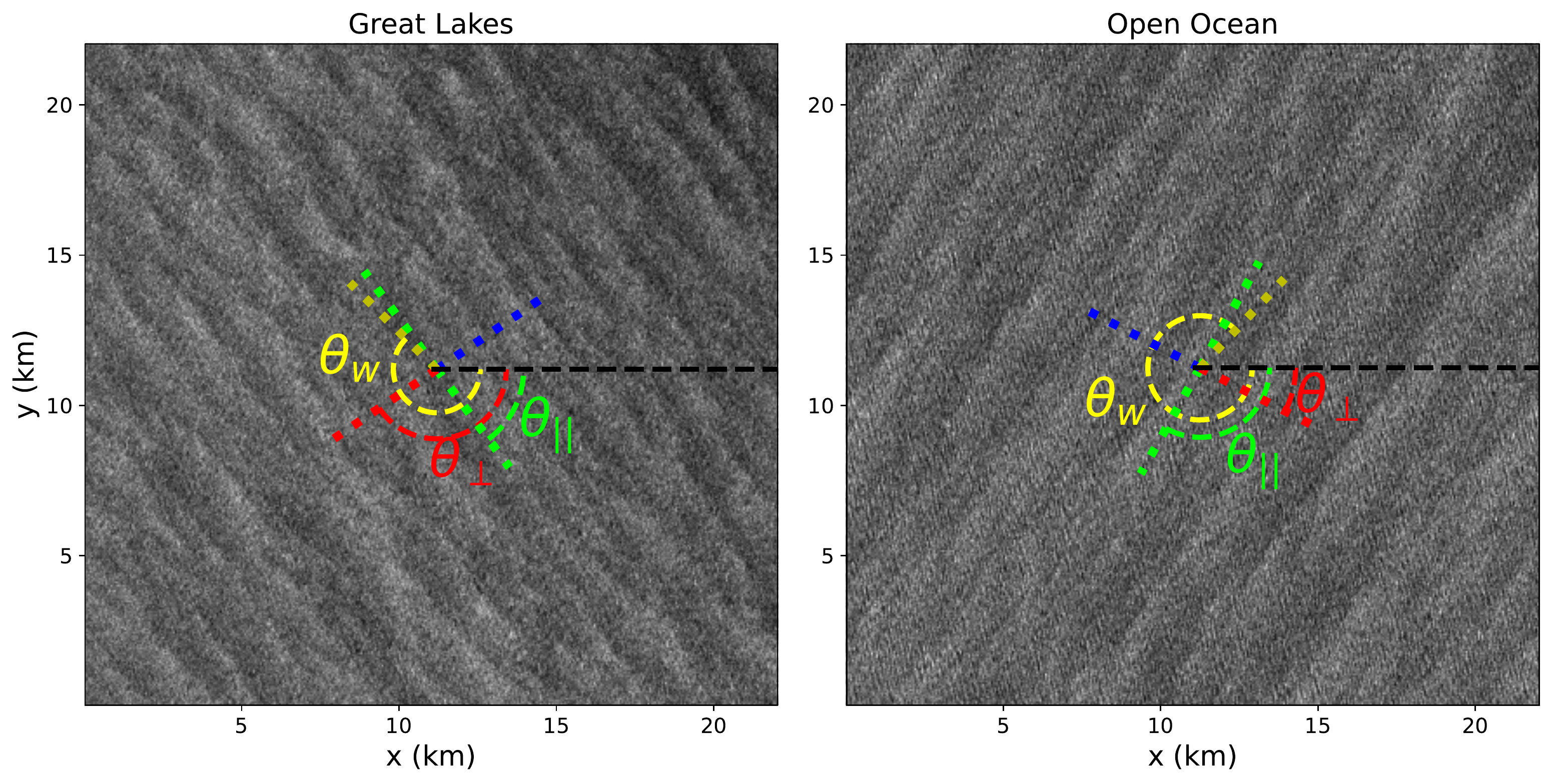}}
\caption{Sentinel-1 SAR images of the ocean: left Great Lakes and right North Pacific. Green dashed lines correspond to the parallel ($\theta_{||}$) and blue and red dashed lines to the perpendicular ($\theta_{\perp}$) directions of convective rolls as obtained from $S_{2}^{l_x,l_y}$. Red indicates the direction with positive skewness and blue with negative one. Yellow dashed line indicates the wind direction ($\theta_{w}$) obtained from the European Centre for Medium-Range Weather Forecasts (ECMWF) global model.}
\label{figImages}
\end{figure}

We work on previously preprocessed Sentinel-1 SAR WV images from https:\/\/xwaves.ifremer.fr\/\#\/~\cite{Wang2020}. These images are corrected from the mean decrease (due to local incidence angle change across the image) of the backscattered signal measured by SAR for a scene assumed homogeneous and present a final spatial resolution after preprocessing of $50$ m which allows us to study convective rolls but also smaller scale events such as swell (typically 200~m; up to 800~m). To illustrate the potential impact of swell on the OLE signature in SAR images, we study here two different images: both obtained by Sentinel-1 A, one between 11:30:04 and 11:30:07 on 2017\/03\/15 from Great Lakes in the border between U.S.A and Canada and another between 20:25:22 and 20:25:24 on 2016\/12\/29 from open North Pacific ocean. While the last one present a significant swell event, the first one is free of this phenomena, see Fig.~\ref{figImages}.

\section{Results and discussion}

\subsection{Second-order structure functions}

\begin{figure}[htbp]
\centerline{\includegraphics[width=\linewidth]{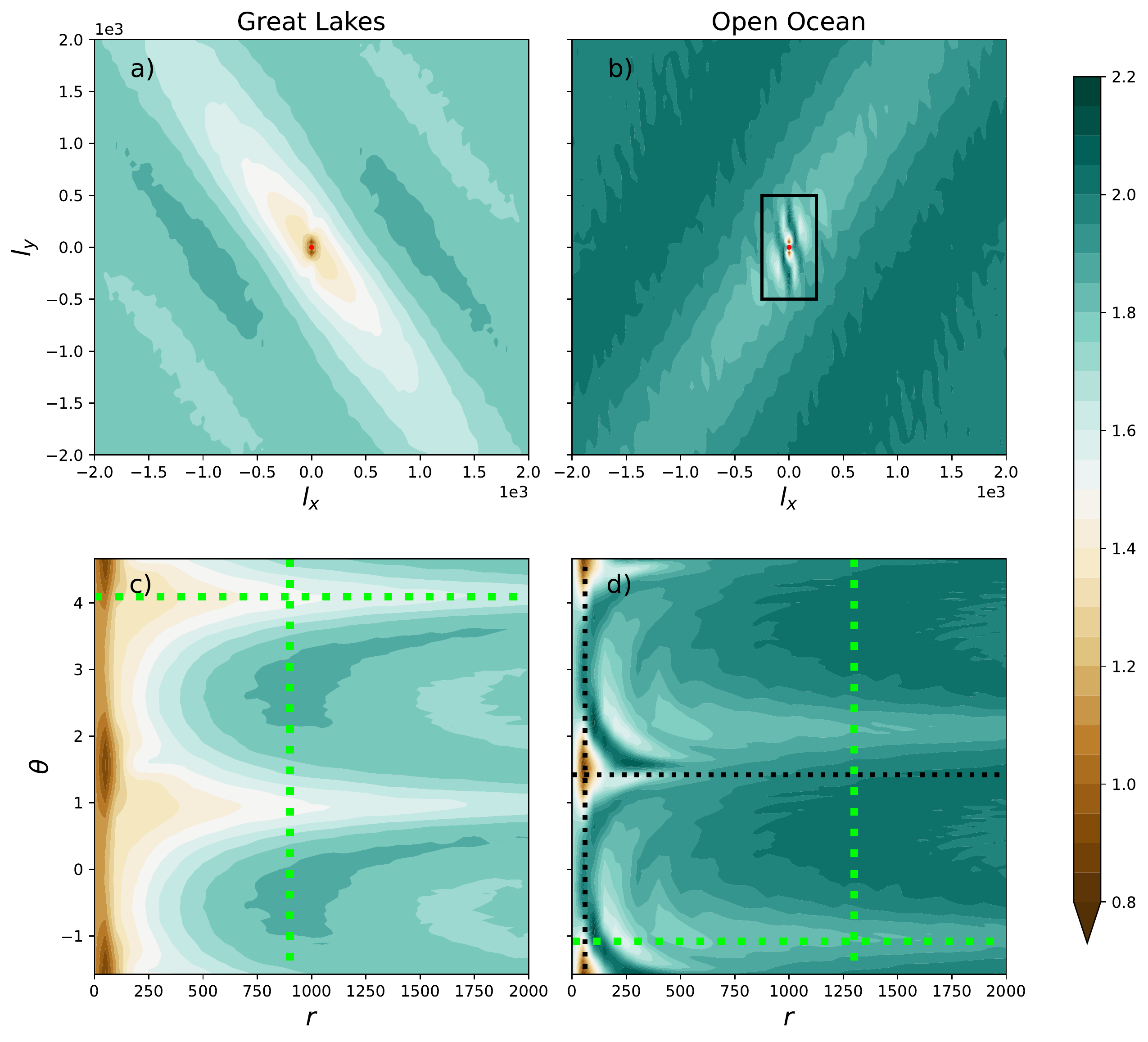}}
\caption{Two-dimensional second-order structure function of two Sentinel-1 SAR images of the ocean: left Great Lakes and right North Pacific. Top in cartesian coordinates $S_{2}^{l_x,l_y}$, bottom in polar coordinates $S_{2}^{r,\theta}$. In a) and b) the center $l_x=l_y=0$ is indicated with a red dot. In figure b) a black square has been added to point out the signature of the swell. In figures c) and d) horizontal and vertical green dashed lines indicate respectively the direction and size of convective rolls. In figure d) black dashed lines indicate the direction and size of swell.}
\label{fig2dS2}
\end{figure}

Fig.~\ref{fig2dS2} shows $S_{2}^{l_x,l_y}$ and $S_{2}^{r,\theta}$ for the Great Lakes and the North Pacific Ocean SAR images presented in section~\ref{sec:data}. For both SAR images, we observe a ellipsoidal structure of a given width about 1Km (green dashed vertical line in polar coordinates) and with its long axis along a given direction $\theta_{||}$ (horizontal green dashed line in polar coordinates). Thus, the second order structure function applied on SAR images characterizes the size and direction of convective rolls, see table~\ref{table:sizedirection}. Same results were obtained with power spectrum~\cite{Wang2020} and correlation~\cite{Lohou1998} methodologies. Moreover, the swell --- if present --- can be characterized in the same way, but looking at smaller scales (see Fig.~\ref{fig2dS2} b). Its presence is revealed by ellipsoidal structures of $S_{2}^{l_x,l_y}$ at small scales (inside the black square in Fig.~\ref{fig2dS2} b). The direction and size of the swell is also obtained (and represented as black dashed lines in Fig.~\ref{fig2dS2} d). We obtain a swell size of 120m in agreement with the WaveWatch3 model (130m) and Fourier analysis (118m). However, the resolution of our measures is the pixel size, in this case 50m, and so swell scales are too small for our method to be adapted.

\begin{table}[htbp]
\caption{Size and direction of convective rolls from Great Lakes and Open ocean SAR images as obtained from $S_{2}^{r,\theta}$.}
\begin{center}
\begin{tabular}{|c|c|c|}
\hline
                        & \textbf{Great Lakes} & \textbf{Open Ocean}   \\ \hline
$\theta_{||}$ (rads)    & $0.95$ ($0.95-\pi$)  & $2.06$ ($2.06-\pi$)   \\ \hline
$\theta_{\perp}$ (rads) & $0.95\pm\pi/2$       & $2.06\pm\pi/2$        \\ \hline
Size (m)                & $1800$               & $2600$                \\ \hline
\end{tabular}
\label{table:sizedirection}
\end{center}
\end{table}

\subsection{Third-order structure functions}

\begin{figure}[htbp]
\centerline{\includegraphics[width=\linewidth]{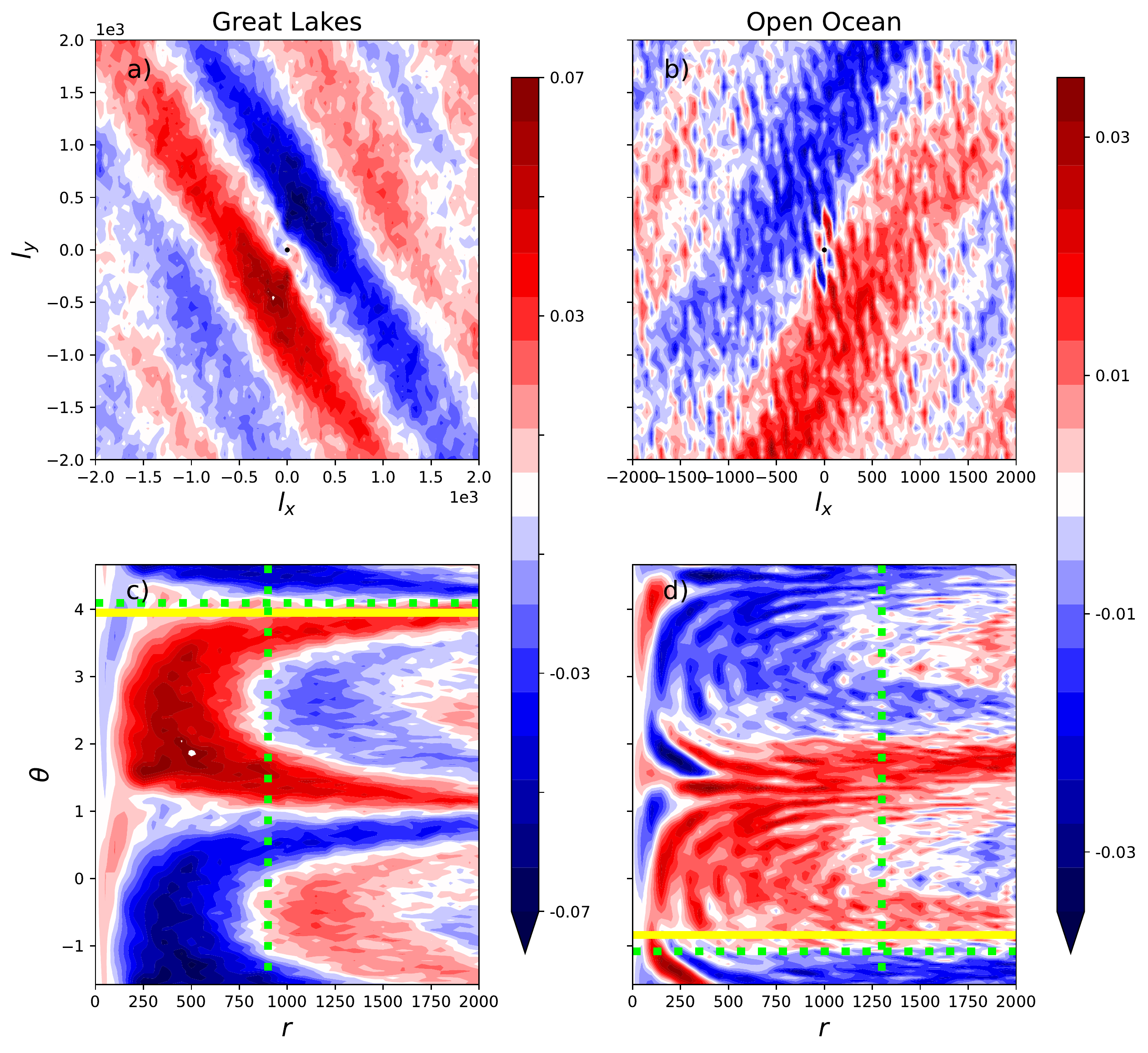}}
\caption{Two-dimensional Skewness across scales of two Sentinel-1 SAR images of the ocean: left Great Lakes and right North Pacific. Top in cartesian coordinates $\mathcal{S}^{l_x,l_y}$, bottom in polar coordinates $\mathcal{S}^{r,\theta}$. In a) and b) the center $l_x=l_y=0$ is indicated with a black dot. In figures c) and d) horizontal and vertical green dashed lines indicate respectively the direction and size of convective rolls and the horizontal yellow line indicates the wind direction.}
\label{fig2dS3}
\end{figure}

Fig.~\ref{fig2dS3} shows $\mathcal{S}^{l_x,l_y}$ and $\mathcal{S}^{r,\theta}$ for the two SAR images. Whereas the line parallel to the direction of the rolls $\theta_{||}$ passing by the center was a symmetry line for $S_{2}^{l_x,l_y}$, it is now an anti-symmetry line for the skewness $\mathcal{S}^{l_x,l_y}$, because $\mathcal{S}^{l_x,l_y}$ is  now an even function of $(l_x,l_y)$: $\mathcal{S}^{l_x,l_y}=-\mathcal{S}^{-l_x,-l_y}$.

For both SAR images, Fig.~\ref{fig2dS3} shows that along $\theta_{||}$ there is no asymmetry, while along the direction perpendicular to the rolls, $\theta_{\perp}$, $\mathcal{S}^{l_x,l_y}\neq0$ for scales smaller than the rolls scale, and so rolls are asymmetrical along this direction, \textit{i.e.} rises and falls of the field are not symmetrical. 
Furthermore, along $\theta_{\perp}$ for scales larger than the roll scales $\mathcal{S}^{l_x,l_y}\approx0$.
Finally, for both images wind direction presents a negative skewness (see yellow lines in Fig.~\ref{fig2dS3} and Fig.\ref{figImages}), pointing out a possible impact of wind direction on the rolls asymmetry.

In Fig.~\ref{fig2dS3} a) strong negative values of $\mathcal{S}^{l_x,l_y}$ are observed when going through $\theta_{\perp}=5.66$ rads. This direction corresponds to the blue dashed line in figure~\ref{figImages} a). Thus $\mathcal{S}^{l_x,l_y}<0$ at scales smaller than the rolls scale and goes to $0$ once this scale is reached. This implies an asymmetry in rolls shape along their perpendicular direction. More precisely, along $\theta_{\perp}=5.66$ rads SAR intensity falls are sharpener than rises and so, we leave rolls more smoothly than we enter into them.
This asymmetry is also observed in $\mathcal{S}^{r,\theta}$, Fig.~\ref{fig2dS3} c).

In Fig.~\ref{fig2dS3} b) and d) negative skewness along $\theta_{\perp}=3.63$ rads indicate also an asymmetry in rolls shape along this direction which corresponds to the blue dashed line in figure~\ref{figImages} b). Along this direction we enter into rolls more sharply than we leave them. Moreover, the swell presence in the SAR image from open ocean adds a small scale effect into the analysis of high-order statistics since swell presents also asymmetries which are characterized by the skewness.

\subsection{Fourth-order structure functions}

\begin{figure}[htbp]
\centerline{\includegraphics[width=\linewidth]{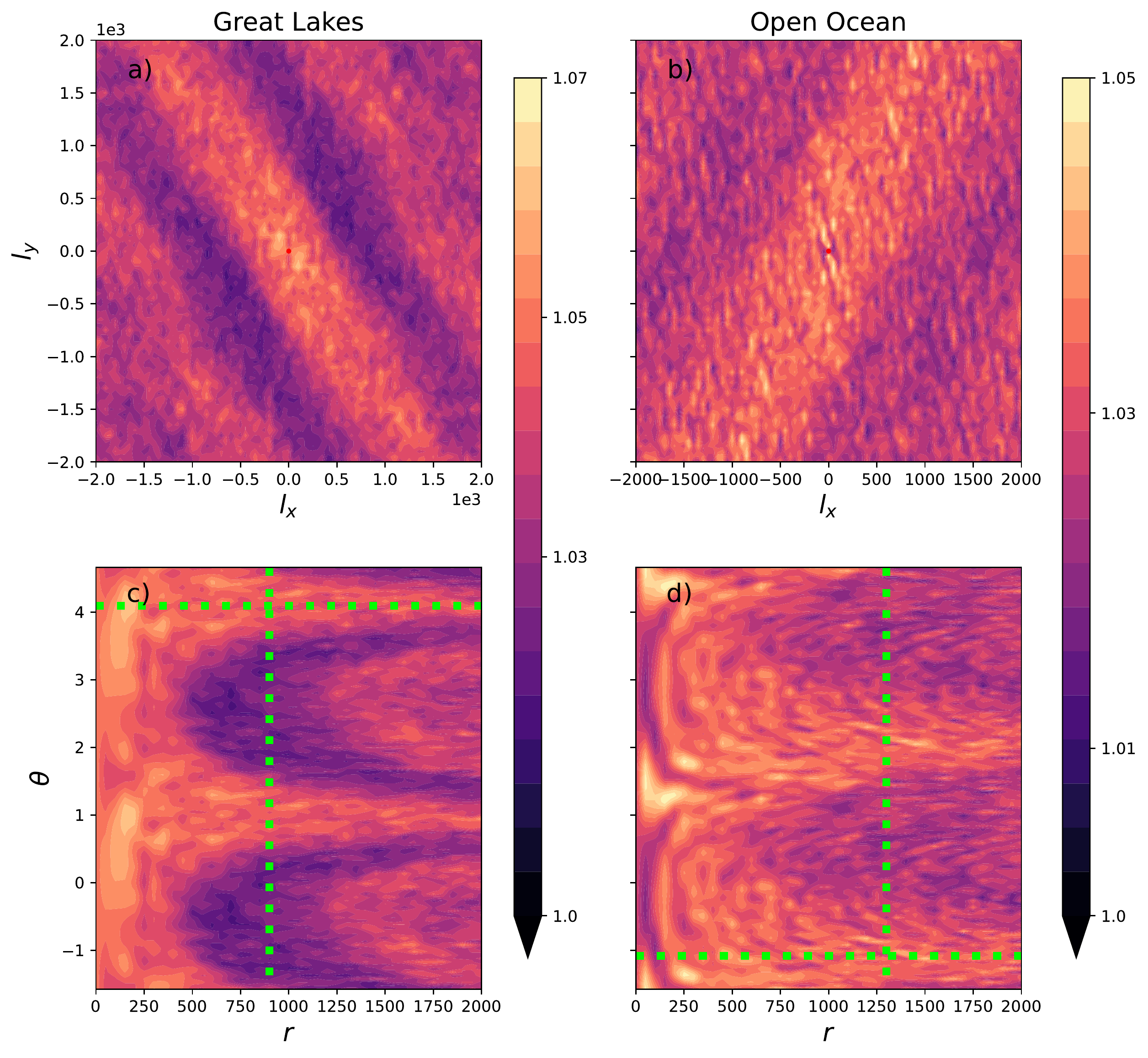}}
\caption{Two-dimensional Flatness across scales of two Sentinel-1 SAR images of the ocean: left Great Lakes and right North Pacific. Top in cartesian coordinates $\mathcal{F}^{l_x,l_y}/3$, bottom in polar coordinates $\mathcal{F}^{r,\theta}/3$. In a) and b) the center $l_x=l_y=0$ is indicated with a red dot. In figures c) and d) horizontal and vertical green dashed lines indicate respectively the direction and size of convective rolls.}
\label{fig2dS4}
\end{figure}

Fig.~\ref{fig2dS4} shows $\mathcal{F}^{l_x,l_y}/3$ and $\mathcal{F}^{r,\theta}/3$ along scales and directions. We observe ellipsoidal structures from which the direction and shape of rolls can be obtained, especially for the Great Lakes case without swell, while for the open ocean case the observation is more blurry. The evolution of $\mathcal{F}^{r,\theta_{\perp}}/3$ across scales indicates the intermittent nature of the rolls evolution along this direction. On the contrary, $\mathcal{F}^{r,\theta_{||}}/3$ seems to remain constant. Finally, the obtained values of the flatness are slightly larger than those of a Gaussian distribution, so implying increment distributions with larger tails than Gaussian.

\subsection{Transects}

\begin{figure}[htbp]
\centerline{\includegraphics[width=\linewidth]{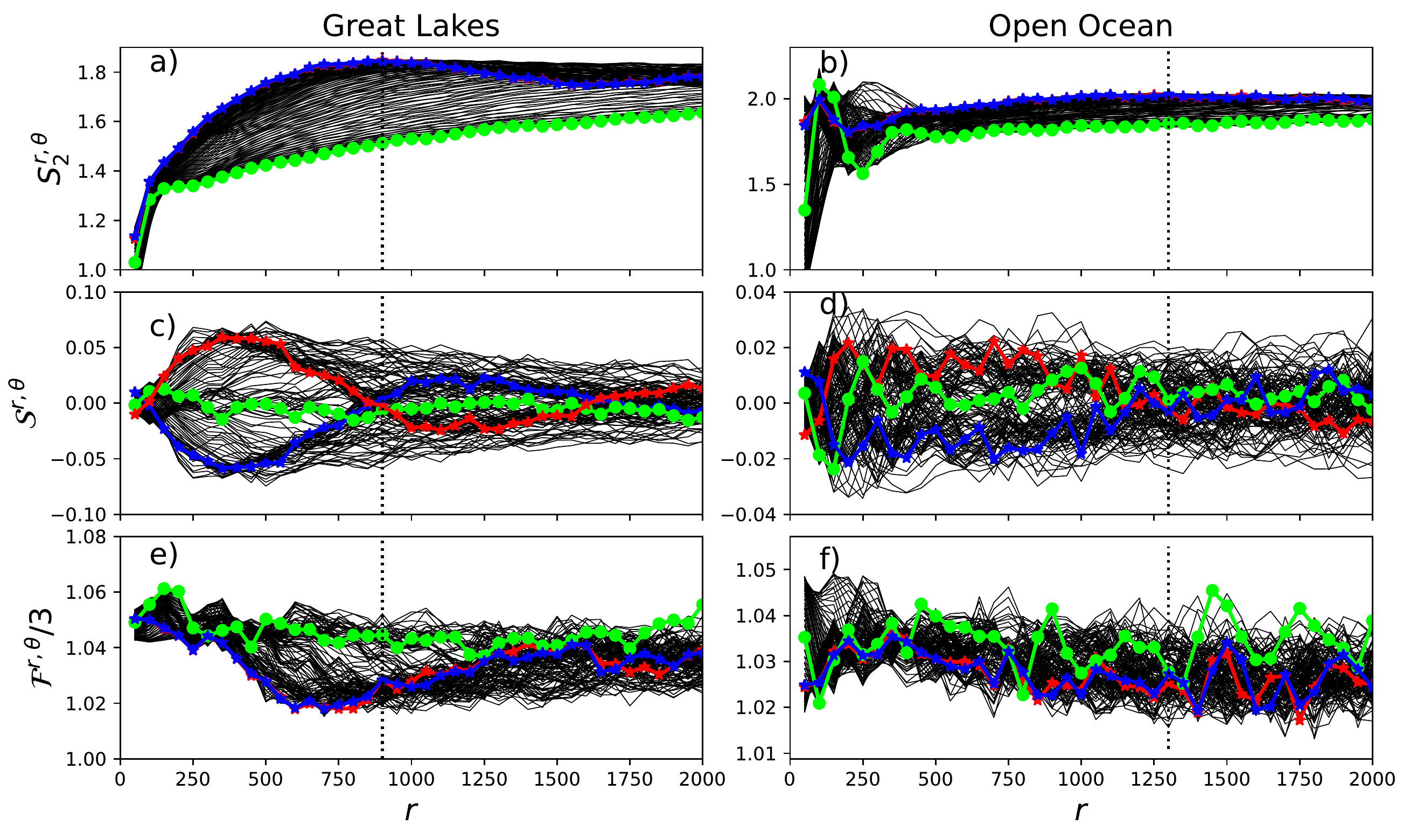}}
\caption{One-dimensional $\theta$-transects of $S_{2}^{r,\theta}$ (a and b), $\mathcal{S}^{r,\theta}$ (c and d) and $\mathcal{F}^{r,\theta}/3$ (e and f) as a function of $r$ for two Sentinel-1 SAR images of the ocean: left Great Lakes and right Open ocean. Direction $\theta_{||}$ is highlighted in green dots and directions $\theta_{\perp}$ and $\theta_{\perp}+\pi$ are highlighted in blue and red asterisks respectively. Vertical black dashed line correspond to the maximum value of $S_{2}^{r,\theta}$ which is used to define the size of the roll.}
\label{fig1dS234}
\end{figure}

Fig.~\ref{fig1dS234} shows $\theta$-transects of $S_{2}^{r,\theta}$, $\mathcal{S}^{r,\theta}$ and $\mathcal{F}^{r,\theta}/3$ as a function of $r$ for both studied SAR images. Black lines correspond to all the studied directions $\theta$, while the direction parallel to the rolls is highlighted in green cercles and the directions perpendicular to the rolls are highlighted in red and blue asterisks. We focus then on $\theta_{\perp}$ and $\theta_{||}$ since they show up the more extreme behaviors with all the other directions in between. 

In Fig.~\ref{fig1dS234} a) and b) $S_{2}^{r,\theta}$ increases from small values at small scale to a higher value plateau at large scale. This plateau is reached faster in the perpendicular direction than along the parallel direction.

Fig.~\ref{fig1dS234} c) and d) show $\mathcal{S}^{r,\theta}$ which evolves very differently depending on $\theta$. Along $\theta_{||}$ $\mathcal{S}^{r,\theta_{||}}$ remains constant and equal to zero. On the contrary, along $\theta_{\perp}$ it is different than zero for scales smaller than the rolls scales and go back to zero for scales larger than the rolls scale. Each perpendicular direction correspond to a different sign of the skewness due to the inverse symmetry explained above. In the case of Great lakes, where swell is not present, $\mathcal{S}^{r,\theta_{\perp}}$ shows the maximum/minimum values. In the case of the open ocean image this is less clear due to the noise induced by the swell.

Finally, Fig.~\ref{fig1dS234} e) and f) show the evolution of $\mathcal{F}^{r,\theta}/3$ for both studied images. 
On the one hand, for Great Lakes $\mathcal{F}^{r,\theta_{||}}/3$ is almost constant at a value close to $1.05$, indicating a distribution slightly more tailed than a Gaussian. On the contrary, $\mathcal{F}^{r,\theta_{\perp}}/3$ decreases and approaches $1$ at scales smaller than the rolls scale. Then at large scales, it turns up to the $\mathcal{F}^{r,\theta_{||}}/3$ value. This evolution of $\mathcal{F}^{r,\theta_{\perp}}/3$ across scales is a signature of intermittency. 
On the other hand, for open ocean $\mathcal{F}^{r,\theta}/3$ seems to be constant for all directions being impossible to discriminate between $\theta_{||}$ and $\theta_{\perp}$. This can be newly due to the swell that difficult the analysis.

\section{Conclusions}

We proposed second and higher order structure functions to analyse the morphology of convective rolls through the study of Sentinel-1 SAR images of the ocean. 
Thus, we showed that $S_{2}^{l_x,l_y}$ characterizes the direction and size of convective rolls, just like correlation~\cite{Lohou1998} or power spectrum~\cite{Wang2020} do.
Moreover, and contrary to correlation and power spectrum, structure functions can be easily generalized ($\mathcal{S}^{l_x,l_y}$ and $\mathcal{F}^{l_x,l_y}/3$) to grasp high-order statistics and so to provide finer information on the shape of convective rolls.
While $\mathcal{S}^{l_x,l_y}$ characterized the asymmetry of convective rolls along the direction perpendicular to the rolls $\theta_{\perp}$, $\mathcal{F}^{l_x,l_y}/3$ pointed out the intermittent nature of convective rolls also along $\theta_{\perp}$. From our knowledge it is the first time that both the rolls asymmetry and intermittency, and consequently their turbulent nature, is described from the analysis of SAR images of the ocean.

In open ocean and on SAR images at 50m resolution, swell can difficult the analysis of high order statistics due to the small scale fluctuations that it produces on the measures. However, preliminary analysis show than a low-pass filtering (or equivalently working on pre-processed SAR images at 200m) could help filtering out the swell signature. Such a filtering may require specific analysis to be adjusted depending on waves properties (e.g. swell orientation with respect to wind direction, swell wavelength ...) 

Future research will deal with better understanding the relationship between rolls asymmetry and atmospheric conditions such as wind direction or shear intensity.
Finally, the codes used in this article to estimate two-dimensional direction-dependent structure functions are provided in open access: \href{https://github.com/cgranerob/2D-Structure-Functions}{https://github.com/cgranerob/2D-Structure-Functions}.

\bibliographystyle{IEEEtran} 
\bibliography{THEBIBLIO}

\end{document}